\documentclass[amsmath,10pt,aps,prl,twocolumn,showpacs,superscriptaddress,groupedaddress]{revtex4-1}%
\usepackage{epsfig,dsfont,amssymb,amsmath,amsthm,amsfonts,amsbsy,mathrsfs}
\usepackage{graphicx}
\usepackage{amsmath}
\usepackage{amssymb}%
\usepackage{bm}
\usepackage{multirow}
\usepackage{times}

\usepackage{color}

\usepackage{hyperref}
\hypersetup{  colorlinks=true, linkcolor=blue, citecolor=red, urlcolor=blue  }

\usepackage{physics}

\theoremstyle{definition}

\begin{document}

\title{Can Chaotic Quantum Circuits Maintain Quantum Supremacy under Noise?}

\author{Man-Hong Yung}
\email{yung@sustc.edu.cn}
\affiliation{Institute for Quantum Science and Engineering and Department of Physics, South University of Science and Technology of China, Shenzhen 518055, China}

\author{Xun Gao}
\email{gaoxun13@mails.tsinghua.edu.cn}   
\affiliation{Center for Quantum Information, Institute for Interdisciplinary Information Sciences, Tsinghua University, Beijing 100084, China} 


\begin{abstract}
Although the emergence of a fully-functional quantum computer may still be far away from today, in the near future, it is possible to have medium-size, special-purpose, quantum devices that can perform computational tasks not efficiently simulable with any classical computer. This status is known as quantum supremacy (or quantum advantage), where one of the promising approaches is through the sampling of chaotic quantum circuits. Sampling of ideal chaotic quantum circuits has been argued to require an exponential time for classical devices. A major question is whether quantum supremacy can be maintained under noise without error correction, as the implementation of fault-tolerance would cost lots of extra qubits and quantum gates. Here we show that, for a family of chaotic quantum circuits subject to Pauli errors, there exists an non-exponential classical algorithm capable of simulating the noisy chaotic quantum circuits with bounded errors. This result represents a serious challenge to a previous result in the literature suggesting the failure of classical devices in simulating noisy chaotic circuits with about 48 qubits and depth 25. Moreover, even though our model does not cover all types of experimental errors, from a practical point view, our result describes a well-defined setting where quantum supremacy can be tested against the challenges of classical rivals, in the context of chaotic quantum circuits. 
\end{abstract}
\maketitle

{\bf Introduction.} A major challenge in the field of quantum computation is related to the question, what exactly are the computational problems quantum computers can solve but classical computers cannot? More precisely, in terms of the language of computational complexity~\cite{kitaev2002classical,nielsen2011quantum}, is the class $\sf BQP$ (bounded-error quantum polynomial time) a proper superset of $\sf BPP$ (bounded-error probabilistic polynomial time) or $\sf P$ (polynomial time)? This question is related to the goal of disproving a fundamental assumption in computer science, known as the extended Church-Turing thesis (ECT)~\cite{nielsen2011quantum}. The ECT thesis, relevant in the asymptotic limit, asserts that, with a polynomial overhead, every physical process (classical or quantum) can be simulated by a classical probabilistic Turing machine. 

Although from the point of view of simulating the dynamics of quantum systems, a quantum computer seems to be capable of exhibiting computational advantages~\cite{Feynman1982,Lloyd1996} over classical devices, there is still no rigorous proof excluding the existence an efficient classical algorithm. Moreover, it is known~\cite{Valiant2001} that not all quantum computational models are hard with classical computation. For example, quantum circuits with only Clifford gates~\cite{Gottesman1998}, with sparse distributions~\cite{Schwarz2013},  fermionic~\cite{Terhal2001,Knill2001a,Jozsa2008} (matchgates), and  bosonic~\cite{Yung2016} (linear optics with finite number of bosons) quantum computation, are classically simulable.

As a step towards this goal, it is important to understand the conditions required to achieve the status of ``quantum supremacy"~\cite{Preskill2012a,Lund2017}, where a quantum device can efficiently perform well-defined tasks beyond the capacities of state-of-the-art of classical computers, even in the absence of quantum error correction. In other words, one has to determine the minimal complexity, in terms of the number of qubits and gates, of a quantum computation model that are not achievable by a classical device. Additionally, one also needs to be sure the status of quantum supremacy of the circuit is robust against the perturbation of experimental noise; in practice, a fault-tolerant implementation of a generic quantum circuit requires an enormous amount of extra resources. By definition, quantum supremacy is a time-dependent concept, depending on the progress of high-performance computing. Practically, the status of quantum supremacy would also built on multiple complexity assumptions or conjectures. 

To achieve quantum supremacy, the computational tasks in question are usually special-purpose, or non-universal, minimizing the technological requirements for a realization. Notably, there are three intermediate models of quantum computation proposed for demonstrating quantum supremacy, including boson sampling~\cite{Aaronson2013a}, one-clean-qubit model (or DQC1) \cite{Knill1998,Morimae2016}, and IQP (instantaneous quantum polynomial-time)~\cite{Bremner2010,Bremner2016a,Bremner2016} and it variants~\cite{Gao2016,Bermejo-Vega2017}. It has been shown~\cite{Bremner2016a,Bremner2016} that a fault-tolerant implementation of boson sampling or IQP sampling is robust against noise in maintaining a quantum advantage. However, without error correction, the distribution of a typical quantum computation can deviate significantly from the ideal implementation, when independent noise is applied to each qubit~\cite{Bremner2016a,Bremner2016}.

Recently, the Google team proposed~\cite{Boixo2016} the use of cross entropy to characterize quantum supremacy. They focused on a computation model based on chaotic quantum circuits, where random quantum gates are applied. There, the authors argued that quantum supremacy is preserved when a certain amount of noise is included. However, it has been shown~\cite{Bremner2016} that IQP circuits, when subject to experimental noise, may become classically simulable and hence lose the status of quantum supremacy. Furthermore, a threshold theorem for quantum supremacy has been established~\cite{Fujii2016}, showing that the threshold value for quantum supremacy is significantly higher than that of universal fault-tolerant quantum computation. This result is consistent with that of Ref.~\cite{Bremner2016}, which showed that a classical algorithm simulating noisy IQP circuits fails after error correction is implemented. However, the applicability of the threshold theorem relies on the assumption of  the ability of performing quantum error correction.  

The central question of interest is ``can chaotic quantum circuits maintain quantum supremacy under noise?", even in the absence of quantum error correction. A major assumption made in Ref.~\cite{Boixo2016} is that, in terms of the cross entropy, the correlation between a family of chaotic quantum circuits and polynomial-time classical algorithms is negligible. The authors in Ref.~\cite{Boixo2016} managed to find a bayesian-based classical algorithm exhibiting the desired behavior, i.e., with an exponential runtime. However, as we shall argue, there exist non-exponential classical algorithms for a family of noisy chaotic circuits. Therefore, the proposal of Ref.~\cite{Boixo2016} suggesting the demonstration of quantum supremacy with 48 qubits becomes questionable in our setting.

To address the open question, here we consider a family of chaotic quantum circuits, where noisy Pauli-$X$ and Pauli-$Z$ gates are probabilistically applied to simulate experimental noises. The goal is to look for non-exponential classical algorithms for simulating the outputs of these chaotic circuits. 

Our results indicate that (i) in order the bound the average $l_1$ norm of these chaotic circuits, a classical algorithm can run in a polynomial time only. (ii) In order to bound the cross entropy, the classical algorithm runs at most in quasi-polynomial time. Our results are established under the same assumption made in Ref.~\cite{Boixo2016}, namely the outputs of the $n$-qubit chaotic quantum circuits obey the Porter-Thomas distribution~\cite{Porter1956}, which is a characteristic of quantum chaos. In fact, we can relax the condition to that the second-moment of the output distribution scales as $O(1/2^n)$. A summary of our main results is included after we introduce all the necessary background. 

However, we do not claim that all chaotic quantum circuits can be simulated efficiently with the proposed classical algorithm. In addition to the need of justifying the assumption mentioned, the family of chaotic circuits considered in this work is a sub-class of chaotic quantum circuits, and the type of experimental noise considered is a toy model. Furthermore, in order to achieve the bounds for a given noise strength $\varepsilon$, there is a factor of $n^{O(1/\varepsilon)}$ in the runtime, which could become a very large number when $\varepsilon \to 0$.

Nevertheless, our model can be taken as a well-defined platform for benchmarking the performance of an experimental implementation of chaotic quantum circuits against the challenge of the non-exponential classical algorithms. At the time of writing, it is possible that as many as 40 to 50 qubits might be already available in research centers from the industry. We propose to test these qubits with random samples of the quantum circuits defined in this work, and compare it with the classical algorithm described below using a reasonably-fast classical workstation. 

On the other hand, we do not claim our classical algorithm to be optimal; there are still rooms to improve the classical algorithm, e.g., by combining compressed sensing or machine learning methods to optimize the random-sampling subroutine. An optimization of our algorithm becomes interesting when its performance can be directly compared with a realistic quantum device.

Finally, perhaps the existence of non-exponential classical algorithms for simulating quantum chaotic processes under noise is by itself an interesting direction for a further investigation, given that classical chaotic systems are notoriously difficult to simulate. Quantum chaotic behaviors include a rapid delocalization of quantum states. Consequently, compared with the overlap of a pair of quantum states generated by Hamiltonians deviated from each other, it decreases exponentially. Therefore, our results suggest that in the presence of noise, the corresponding overlap may not decrease exponentially anymore. This result is consistent in the asymptotic limit where the noise is so strong that the outcomes of all quantum circuits, chaotic or not, become identical to the uniform distribution. 
 

\section{ --- \ Part I: setting the stage \ ---}

{ \bf Three different sources of bit-strings.} In this work, we are considering outputs in the form of $n$-bit strings of $0$'s and $1$'s, i.e., $x \in \{ 0,1\}^n$, coming from three different sources, namely (i) ideal quantum circuits, (ii) noisy quantum circuits modeling experimental implementations, and (iii) randomized classical algorithms. We denote the corresponding probability distributions as, respectively,
\begin{itemize}
  \item[(i)] ${P_{\rm qc}}\left( x \right) = {\left| {\left\langle x \right|U\left| {{\psi _{\rm in}}} \right\rangle } \right|^2}$ \ ,  
  \item[(ii)] ${P_{\exp }}\left( x \right) = \left\langle x \right|\mathcal{E}\left( {\left| {{\psi _{\rm in}}} \right\rangle \left\langle {{\psi _{\rm in}}} \right|} \right)\left| x \right\rangle $ \ ,
  \item[(iii)] ${P_{\rm cl}}\left( x \right)$ \ .
\end{itemize} 

Here the quantum state $\left| {{\psi_{\rm in}}} \right\rangle$ is the initial state of the quantum circuit, the operator~$U$ is the unitary transformation generated by the quantum circuit, and the super-operator $\mathcal{E}(\cdot)$ represent some noisy channel, modeling an experimental implementation. We shall describe the classical algorithm in a later section (see Eq.~(\ref{itlrwdPcl})). 

In fact, instead of a single circuit, we shall consider an ensemble of quantum circuits, where each circuit is labeled by a string $x'$, i.e., $U_{x'}$. Therefore, it is necessary to compare the corresponding (conditional) probabilities for each circuit, i.e., 
\begin{equation}
{P_{{\text{qc}}}}\left( {x|{U_{x'}}} \right), \  {P_{{\text{exp}}}}\left( {x|{U_{x'}}} \right), \ {P_{{\text{cl}}}}\left( {x|{U_{x'}}} \right)  \ .
\end{equation} 

Later, we shall distinguish two types of classical algorithms. The first type aims to obtain the numerical values of ${P_{{\text{cl}}}}\left( {x|{U_{x'}}} \right)$. In the second type, we aim to produce a classical distribution described by ${P_{{\text{cl}}}}\left( {x|{U_{x'}}} \right)$. The relationship between them is rather subtle.

{\bf Vector norm.} In order to characterize the performance of simulation from one source to another, it is common (see e.g. Refs.~\cite{Bremner2016a,Bremner2016}) to consider the $l_1$-norm between two distributions, e.g., 
\begin{equation}\label{L1norm_1stap}
\Lambda \equiv {\left\| {{P_{\exp }} - {P_{{\text{cl}}}}} \right\|_1} = \sum\limits_{x \in {{\left\{ {0,1} \right\}}^n}} {\left| {{P_{\exp }}\left( x \right) - {P_{{\text{cl}}}}\left( x \right)} \right|} \ ,
\end{equation}
In particular, subject to a couple of reasonable conjectures, if there exists a classical algorithm that can sample any IQP (instantaneous quantum polynomial time) circuit to a small constant in the $l_1$ norm, then the polynomial hierarchy collapses to the third level~\cite{Bremner2016a}, which is widely believed to be implausible. 


{\bf Cross entropy.} Alternatively, it has been suggested~\cite{Boixo2016} that one may consider the cross entropy $S_c$ between two distributions, e.g., 
\begin{equation}\label{cross_ent_exp_qc}
{S_c}\left( {{P_{\exp }},{P_{{\text{qc}}}}} \right) \equiv  - \sum\limits_{x \in {{\left\{ {0,1} \right\}}^n}} {{P_{\exp }}\left( x \right)\log } \ {P_{{\text{qc}}}}\left( x \right) \ .
\end{equation}
The operational meaning of the cross entropy is as follows: suppose we generate a sequence of many, $m \gg 1$, independent bit-strings $x$'s from each source, e.g. from an ideal quantum circuit $s_{\rm qc}=\{ x_1^{\rm qc},x_2^{\rm qc},...,x_m^{\rm qc}\}$, and experiment $s_{\rm exp}=\{ x_1^{\rm exp},x_2^{\rm exp},...,x_m^{\rm exp}\}$. The probability of obtaining the sequence $s_{\rm qc}$ is given by the product, 
\begin{equation}
\Pr \left( {{s_{{\text{qc}}}}} \right) = \prod\nolimits_{i = 1}^m {{P_{{\text{qc}}}}\left( {x_i^{{\text{qc}}}} \right)} \ .
\end{equation}

Now, we can also ask the question: how likely does the {\it same} quantum circuit generate the sequence $s_{\rm exp}$ in the experiment? The corresponding probability is given by 
\begin{equation}
\Pr \left( {{s_{\exp }}} \right) = \prod\nolimits_{i = 1}^m {{P_{{\text{qc}}}}\left( {x_i^{\exp }} \right)}\ .
\end{equation}
 (Note that we have to use ${{P_{{\text{qc}}}}}$ instead of ${{P_{{\text{exp}}}}}$.) After applying the central-limit theorem (see e.g. Ref.~\cite{Boixo2016}), the probability $\Pr \left( {{s_{\exp }}} \right)$ can be determined by the cross entropy in Eq.~(\ref{cross_ent_exp_qc}) through, 
\begin{equation}
\Pr \left( {{s_{\exp }}} \right) = {e^{ - m{S_c}\left( {{P_{\exp }},{P_{{\text{qc}}}}} \right)}} \ .
\end{equation}

{\bf Shannon versus cross entropy.} In the noise-free (or ideal) limit, we expect that the two probabilities become the same, i.e., $\Pr \left( {{s_{\exp }}} \right)\to \Pr \left( {{s_{{\text{qc}}}}} \right)$, which means that the cross entropy~$S_c$ becomes identical to the Shannon entropy, 
\begin{equation}
{S}\left( {{P_{{\text{qc}}}}} \right) \equiv  - \sum\nolimits_{x \in {{\left\{ {0,1} \right\}}^n}} {{P_{{\text{qc}}}}\left( x \right)\log } \ {P_{{\text{qc}}}}\left( x \right) \ ,
\end{equation}
i.e., ${S_c}\left( {{P_{\exp }},{P_{{\text{qc}}}}} \right)\to {S}\left( {{P_{{\text{qc}}}}} \right)$. 

More generally, one may quantify the performance of the experiment by the difference of the two quantities, 
\begin{equation}
\Delta_{\rm exp}\equiv \left| {S\left( {{P_{{\text{qc}}}}} \right) - {S_c}\left( {{P_{\exp }},{P_{{\text{qc}}}}} \right)} \right| \ .
\end{equation}
In a similar way, we may consider quantifying the performance of a classical algorithm by
\begin{equation}
{\Delta _{{\text{cl}}}} \equiv \left| {S\left( {{P_{{\text{qc}}}}} \right) - {S_c}\left( {{P_{{\text{cl}}}},{P_{{\text{qc}}}}} \right)} \right| \ .
\end{equation}
 Note that for the purpose of demonstrating quantum supremacy, it is not necessary for ${\Delta_{\rm exp}}$ to be small. Instead, one should make sure ${\Delta _{\rm exp}}$ to be significantly less than ${\Delta _{\rm cl}}$, for all possible classical algorithms.

{\bf Constraint of quantum supremacy.} Now, applying the triangle inequality to ${\Delta _{\rm cl}}$, we have, 
\begin{equation}
{\Delta _{\rm cl}} \ \leqslant \ {\Delta_{\rm exp}}+ \Delta_{S} \ ,
\end{equation}
where 
\begin{equation}\label{DSScScw}
\Delta_{S} \equiv \left| {{S_c}\left( {{P_{\exp }},{P_{{\text{qc}}}}} \right) - {S_c}\left( {{P_{{\text{cl}}}},{P_{{\text{qc}}}}} \right)} \right| 
\end{equation}
is the main quantity we shall focus on. This inequality implies that if there exists an efficient classical algorithm such that ${\Delta _S}$ can be bounded by a small constant, then the performance of the classical algorithm, in terms of simulating the ideal quantum circuit, cannot be much worse than that of the experiment. 

In other words, for the cases where the values of $\Delta_S$ are small, the value of cross entropy cannot be used to justify the status of quantum supremacy in an experimental implementation. Of course, a similar argument is applicable to the case using the $l_1$ norm (see Eq.~(\ref{L1norm_1stap})). 

{\bf Summary of main results.} 
In this work, we ask the following question: given an ensemble of chaotic quantum circuits, how well can classical algorithms approximate  experimental implementations of the circuits? In Ref.~\cite{Boixo2016}, the performance of a classical algorithm simulating the chaotic quantum circuit was exponentially close to the performance of  using a simple uniform distribution. The result gives a lower bound on the performance of classical algorithms, but it does not exclude the possibility of the existence of better classical algorithms, which is the main task of this work.

More precisely, the authors of Ref.~\cite{Boixo2016} employed a measure defined by 
\begin{equation}
{\Delta _H}\left( {{P_{{\text{cl}}}}} \right) \equiv {S_0} - S_c\left( {{P_{{\text{cl}}}},{P_{{\text{qc}}}}} \right) \ ,
\end{equation}
where ${S_0}$ is the cross entropy when a uniform distribution is compared with the quantum circuit. It was shown that if a classical algorithm can perfectly reproduce the distribution of the quantum circuit, i.e., ${P_{{\text{cl}}}} \to {P_{{\text{qc}}}}$, then $\Delta_H \left( {{P_{{\text{qc}}}}} \right)$ becomes unity, i.e., ${\Delta _H}\left( {{P_{{\text{qc}}}}} \right) \to 1$. However, it approaches zero ${\Delta _H} \to 0$, if we replace it with a uniform distribution, i.e., ${P_{{\text{cl}}}}\left( x \right) \to 1/N$.

The central idea of Ref.~\cite{Boixo2016} is that if it was true that for a class of quantum circuits, the value of ${\Delta _H}\left( {{P_{{\text{qc}}}}} \right)$ for all polynomial-time classical algorithms is significantly smaller than that the values ${\Delta _H}\left( {{P_{\exp }}} \right)$ resulting from experimental implementations, then the status of quantum supremacy can be achieved. In our context, one needs to show that 
\begin{equation}
{\Delta _S} = \left| {{\Delta _H}\left( {{P_{\exp }}} \right) - {\Delta _H}\left( {{P_{{\text{cl}}}}} \right)} \right| \ ,
\end{equation}
to be very different from zero, comparing with experimental implementation and the best classical algorithm.

Here we consider a class of chaotic quantum circuits, where the imperfection in their experimental implementations are modeled by random bit-flip errors and depolarizing noise. Our goal in this work is to show that there exists (quasi) polynomial-time classical algorithms that can simulate a class of chaotic quantum circuits under noise. In other words, our results indicate that in some situations, it can happen that a polynomial-time classical algorithm leads to a small ${\Delta _S}$.

More precisely, our classical algorithm can output the values of the classical probabilites such that, on average, the $l_1$ norm with the experimental simulation ${\left\| {{P_{\exp }} - {P_{{\text{cl}}}}} \right\|_1}$ is bounded by a given small constant $\delta$. The runtime of the classical algorithm scales as $O( {{{(n + m)}^L}/{\delta ^2}} )$, where $n$ is the number of qubits in the quantum circuits, $m=\text{poly}(n)$ is the number of ancilla qubits. Here $L = O(\log ({\alpha / \delta ^2} )/\varepsilon )$, where $\varepsilon$ is the strength of the noise and $\alpha =O(1)$.

The classical algorithm can then be employed to produce a classical distribution of bit strings, approximating those from the experimental simulation. In particular, if we randomly choose a quantum circuit in our ensemble, with a probability at least $1-2/k^2$ for any $k>1$, the $l_1$ norm between the probability distribution generated by the classical algorithm (denoted as ``$\text{Alg}$") and that of experimental simulation is bounded by ${\left\| {{\text{Alg}} - {P_{{\text{exp}}}}} \right\|_1} \leqslant 4k\delta /\left( {1 - k\delta } \right)$. 

Furthermore, on average, the difference in cross entropy in Eq.~(\ref{DSScScw}) is bound by $O( {\delta \sqrt {{{\left( {n\log 2 + \gamma } \right)}^2} + {\pi ^2}/6} })$, where $\gamma = 0.57721 ...$ is the Euler constant. In Ref.~\cite{Boixo2016}, $n=48$ is taken to be the ``supremacy frontier". 

The assumption involved in this work is the chaotic nature of the ensemble of the quantum circuits. Starting from a chaotic quantum circuit of $n$ qubits (to be defined later), the other members in the ensemble are generated by applying single-qubit Pauli-$X$  gates to each of the local gates in the original circuit. We assume these quantum circuits remains chaotic.

{\bf Overview on technical details.} In the remaining of this work, we shall first define chaotic quantum circuits, which involves an requirement that the second moment of the probability distribution be bounded by $O(1/2^n)$. {The main idea is inspired by the techniques in Ref.~\cite{Bremner2016} and Ref.~\cite{Gao2016}}. In addition, we start with a circuit decomposition of a chaotic quantum circuit based in the universal set 
\begin{equation}
\left\{ {H,Z\left( \alpha  \right),CZ} \right\} \ .
\end{equation}
In this way, we are able to encode the original circuit as one of the branches in a IQP circuit {using the idea in Ref.~\cite{Gao2016} which maps an ensemble of random circuit to an IQP circuit}. Each circuit in the other branches represent a variant of the original circuit subject to some additional Pauli-$X$ gates.

To model noise, a depolarizing channel is applied to each of the qubits in the IQP circuit. In this way, we show that effectively each of the local gates are applied with a bit-flip error in the chaotic quantum circuits. Further analysis are performed by including several results in Ref.~\cite{Bremner2016}, which proposed the use of Fourier components in IQP circuits to simplify calculations of probability distributions. In particular, it is shown that the effect of including depolarizing noise on IQP circuits is to reduce the size of the Fourier components, which makes it possible to approximate the probability distribution through truncation of the small Fourier components. 

Finally, we found that the chaotic nature of the quantum circuits implies that the difference in the cross entropy can be bounded in a similar way as the $l_1$ norm. However, bounding the cross entropy difference to a small constant is more demanding than that for the $l_1$ norm. The former case requires a quasi-polynomial time algorithm. 

Before we get started on the technical details, we remark that most of the technical work presented here is devoted to provide a rigorous proof on the lower bound of the performance of the proposed classical algorithms. In practice, we expect the average performance should be better than the lower bound presented here.

Furthermore, one may consider performing additional (including heuristic) optimization through other  classical methods to further improve the efficiency. For example,  one may apply the technique of compressed sensing to reduce the number of sampling on the Fourier components to reduce computational costs. Therefore, the classical algorithms outlined in this work provide a framework for comparing the performance of near-future experimental demonstrations of quantum supremacy based on chaotic quantum circuits.
 
\section{--- \ Part II: Chaotic Quantum circuits \ ---}
{\bf Chaotic quantum circuits.} In this work, we are interested in some chaotic quantum circuit~$U_0$ applied to a fixed initial state, namely 
\begin{equation}\label{input_nqs}
 \ket{\psi_{\rm in}} \equiv {\left|  +  \right\rangle ^{ \otimes n}} = {H^{ \otimes n}}\left| {000...0} \right\rangle \ ,
\end{equation}
 of $n$ qubits, where $H \equiv \left|  +  \right\rangle \left\langle 0 \right| + \left|  -  \right\rangle \left\langle 1 \right|$ is the Hadamard gate and $\left|  \pm  \right\rangle  \equiv \left( {\left| 0 \right\rangle  \pm \left| 1 \right\rangle } \right)/\sqrt 2 $. Here chaotic quantum circuits~\cite{Hangleiter2017} correspond to the class of quantum circuits where the values of 
\begin{equation}
{P_{\rm qc}(x | U_0)} \equiv {\left| {\left\langle x \right|{U_0}\left| {{\psi _{{\text{in}}}}} \right\rangle } \right|^2}
\end{equation} 
obey the Porter-Thomas distribution (i.e., the distribution of the distribution), $N^2 \exp \left( { - NP} \right)$ for $N=2^n$. In fact, we will only need a weaker condition that the second moment is $O(1/N)$, i.e., 
\begin{equation}\label{2ndm_la2n}
{R_0} \equiv {\sum\limits_{x \in {{\left\{ {0,1} \right\}}^n}} {{P_{{\text{qc}}}}\left( {x|{U_0}} \right)} ^2} \ \leqslant \ \frac{{{\alpha _0}}}{2^n} \ ,
\end{equation}
for some constant $\alpha_0$. Note that for quantum circuits obeying strictly the Porter-Thomas distribution, 
\begin{equation}
\left\langle {{P^2}} \right\rangle  \equiv N^2\int_0^\infty  {{e^{ - NP}}{P^2}dP = } 2/{N} \ ,
\end{equation}
implying that $Q_0= 2/2^n$.

{\bf Circuit decomposition.} In addition, we consider the chaotic quantum circuit $U_0$  described by a universal gate set $\left\{ {H,Z\left( \alpha  \right),CZ} \right\}$, where $Z\left( \alpha  \right) \equiv \left| 0 \right\rangle \left\langle 0 \right| + {e^{i\alpha }}\left| 1 \right\rangle \left\langle 1 \right|$ covers all possible values of~$\alpha$, and $CZ \equiv I - 2\left| {11} \right\rangle \left\langle {11} \right|$. We further require that each single-qubit gate is always realized by the product
\begin{equation}
J\left( \alpha  \right) \equiv HZ\left( \alpha  \right) \ ,
\end{equation}
which is possible because any single qubit unitary gate $U_\text{qubit}$ can always be decomposed into at most four applications of $J(\alpha)$, i.e., ${U_{{\text{qubit}}}} = {e^{i\xi }}J\left( 0 \right)J\left( \alpha  \right)J\left( \beta  \right)J\left( \gamma  \right) = {e^{i\xi }}Z\left( \alpha  \right)X\left( \beta  \right)Z\left( \gamma  \right)$.

Consequently, such a chaotic quantum circuit can be generated by a post-selection in the setting of measurement-based quantum computation. In particular, for a general qubit state~$\left| \psi  \right\rangle$ and another qubit in state $\ket{+}$, when we apply a $CZ$ gate to correlate them followed by applying $J\left( \alpha  \right) \equiv HZ\left( \alpha  \right)$ to the first qubit, we obtain the following state before measurement (after a swapping the two qubits for clarity),
\begin{equation}\label{MBQC_0J1X}
 \left| {{\psi _J}} \right\rangle = \frac{1}{{\sqrt 2 }}  ( J\left( \alpha  \right)\left| \psi  \right\rangle  \otimes \left| 0 \right\rangle  + XJ\left( \alpha  \right)\left| \psi  \right\rangle \otimes \left| 1 \right\rangle    ) \ ,
\end{equation}
where $\left| {{\psi _J}} \right\rangle  \equiv ({J}\left( \alpha  \right)\otimes I) \ CZ\left( {\left| \psi  \right\rangle \left|  +  \right\rangle } \right)$ and $X \equiv \left| 1 \right\rangle \left\langle 0 \right| + \left| 0 \right\rangle \left\langle 1 \right|$. Hence, for example, the $J(\alpha)$ gate is applied if the state $\ket{0}$ is obtained in a measurement. 

{\bf Ensemble from chaotic quantum circuits.} However, we are not interested in physically generating such a quantum circuit by post-selection, which is highly inefficient when scaled up. Instead, the purpose here is to take the chaotic quantum circuit $U_0$ as the ``seed" for describing an ensemble of quantum circuits generated by applying bit-flip gates after each $J$-gate in the original quantum circuit $U_0$. For example in Eq.~(\ref{MBQC_0J1X}), the  quantum circuit obtained by the result $\ket{1}$ is the one with a Puali-$X$ gate (or bit-flip) applied after the $J(\alpha)$ gate. 

Generally, for a quantum circuit $U_0$ with $m$ single-qubit $J$ gates, there are exactly $2^m$ different quantum circuits in the ensemble. We label each quantum circuit $U_{x'}$ by a $m$-bit string~
\begin{equation}
x'={x'_1} \ {x'_2} \ {x'_3} \cdots \  {x'_m} \ .
\end{equation}
The original one is always $x'=000 \cdots 0$, i.e., $U_0 = U_{000 \cdots 0}$. Furthermore, if the $j$-th bit ($x'_j$) is non-zero, it means that a Pauli-$X$ gate is applied to the $j$-th $J$ gate in $U_0$.

{\bf Robustness of chaotic circuits.} The only assumption behind this work is that  an application of these Pauli-$X$ gates does not change the chaotic nature of the original quantum circuit $U_0$, at least on average. More specifically, given the condition in Eq.~(\ref{2ndm_la2n}), we assume it still holds that for some constant $\alpha$,
\begin{equation}\label{assum_Q2n}
\frac{1}{{{2^m}}}\sum\limits_{{x'} \in {{\left\{ {0,1} \right\}}^m}} {{R_{x'}}}  \leqslant \frac{\alpha }{{{2^n}}} \ ,
\end{equation}
where ${R_{x'}} \equiv \sum\nolimits_{x \in {{\left\{ {0,1} \right\}}^n}} {{P_{{\text{qc}}}}{{\left( {x|{U_{x'}}} \right)}^2}}$ is the second moment of circuit $U_{x'}$. In fact, we shall see that such an assumption is equivalent to the assumption made in Ref.~\cite{Bremner2016} for the case of random IQP circuit. Furthermore, a similar but different approach of generating random ensemble of quantum circuits have been proposed~\cite{Bermejo-Vega2017}, where the last set of Hamdamard gates are not applied to the ancilla qubits before measurement. There, numerical evidence was provided for showing the convergence of the probability distribution towards the Porter-Thomas distribution. For the sake of argument, here we keep the condition in Eq.~(\ref{assum_Q2n}) as an assumption.

\section{--- \ Part III: Universal IQP circuits\ ---}

{\bf Encoding unitaries in IQP circuit.} 
Now, we can extend the argument in Eq.~(\ref{MBQC_0J1X}) to the whole quantum circuit of $U_0$, in the usual way described in measurement-based quantum computation. However, the difference is that it is sufficient to consider the procedure in an non-adaptive manner as follows. 

Suppose there are $m+n$ qubits initialized in the $\ket{0}$ state. Then, we apply Hadamard gates to all the qubits, i.e., ${H^{ \otimes (m+n )}}\left| {{0^{m+n}}} \right\rangle $. Next, we apply a diagonal gate $D$, which contains three sets of commuting gates: (i) all the $CZ$ gates in the original quantum circuit $U_0$ and (ii) the $CZ$ gates with the $m$ ancilla qubits generating the $J$ gates, (iii) the $Z$ gates associated with each $J$ gate. For example, the case of Eq.~(\ref{MBQC_0J1X}) corresponds to the case $D = ( Z(\alpha) \otimes I ) \ CZ $, with $\left| \psi  \right\rangle  = \left| 0 \right\rangle$.

Finally, when this unitary gate, 
\begin{equation}
U_{\rm IQP}\equiv H^{ \otimes (m+n )} \, D \, H^{ \otimes (m+n )} \ .
\end{equation}
is applied to the initial state $\left| {{0^{n + m}}} \right\rangle$ (without loss of generality), we have 
\begin{equation}\label{encode_IQP_QC}
{U_{{\text{IQP}}}}\left| {{0^{n + m}}} \right\rangle  = \frac{1}{{{2^{m/2}}}} \sum\limits_{x' \in {{\left\{ {0,1} \right\}}^m}} {U_{x'}}\left| {{\psi_{\rm in}}} \right\rangle \otimes  {\left| x' \right\rangle }  \ ,
\end{equation}
where $\left| {{\psi _{{\text{in}}}}} \right\rangle$ is defined in Eq.~(\ref{input_nqs}), and $U_{x'}$ is the Pauli-$X$ variants of the original quantum circuit $U_0$ discussed previously. 

It is now evident that any member $U_{x'}$ in the ensemble of chaotic quantum circuits is an instance after a partial measurement of a IQP circuit. Of course, one may deterministically pick any choice of the quantum circuit $U_{x'}$ for an experimental implementation. 

Note that we have the following relation: 
\begin{equation}
{P_{{\text{IQP}}}}\left( {x,x'} \right) = {P_{{\text{qc}}}}\left( {x|{U_{x'}}} \right)P\left( {x'} \right) \ ,
\end{equation}
or explicitly,
\begin{equation}\label{PIQP2mxuin}
{P_{{\text{IQP}}}}\left( {x,x'} \right) = \frac{1}{{{2^m}}}{\left| {\left\langle x \right|{U_{x'}}\left| {{\psi _{{\text{in}}}}} \right\rangle } \right|^2} = \frac{1}{{{2^m}}}{P_{{\text{qc}}}}\left( {x|{U_{x'}}} \right) \ ,
\end{equation}
since $P\left( {x'} \right) = 1/{2^m}$ and ${P_{{\text{qc}}}}\left( {x|{U_{x'}}} \right) = {\left| {\left\langle x \right|{U_{x'}}\left| {{\psi _{{\text{in}}}}} \right\rangle } \right|^2}$.

{\bf Hadamard transform and IQP circuits.}
Note that the elements of the diagonal gate $D$ can be sorted out readily as a function,
\begin{equation}
f\left( {x,x'} \right) \equiv \left\langle {x,x'} \right|D\left| {x,x'} \right\rangle \ ,
\end{equation}
of the system and ancilla qubits. Given the initial state $\left| {{0^{n + m}}} \right\rangle$, the Hamdamard gates produce a uniform superposition of quantum states, which means that $D{H^{ \otimes (n + m)}}\left| {{0^{n + m}}} \right\rangle  = {2^{ - (n + m)/2}}\sum\nolimits_{x,x'} {f\left( {x,x'} \right)} \left| {x,x'} \right\rangle$. Applying to any computational basis, 
\begin{equation}\label{Hn_atacb}
{H^{ \otimes (n + m)}}\left| {x,x'} \right\rangle  = \sum\limits_{y,y'} {\frac{{{{\left( { - 1} \right)}^{x \cdot y + x' \cdot y'}}}}{{{2^{(n + m)/2}}}}} \left| {y,y'} \right\rangle  ,
\end{equation}
which implies that the probability,
\begin{equation}
P_\text{IQP}\left( {x,x'} \right) \equiv {\left| {\left\langle {x,x'} \right|{U_{{\text{IQP}}}}\left| {{0^{n + m}}} \right\rangle } \right|^2} \ ,
\end{equation}
from the IQP circuit is given by,
\begin{equation}\label{PIQP_fsum}
{P_{{\text{IQP}}}}\left( {x,x'} \right) = \frac{1}{{{2^{n + m}}}}{\left| {\sum\limits_{y,y'} {f\left( {y,y'} \right)} {{\left( { - 1} \right)}^{xx' \cdot yy'}}} \right|^2} \ ,
\end{equation}
where $y \in \{ 0,1\}^n$, $y' \in \{ 0,1 \} ^m$, and $xx' \cdot yy' \equiv x \cdot y + x' \cdot y'$.

{\bf Adding noises to quantum circuits.} Let us now turn our attention to the problem of modeling the noise in experimental implementations of these chaotic quantum circuits. For the sake of argument, we consider adding two types of noises to the quantum circuits $U_{x'}$, namely (i) bit-flip noise $X$ to each single-qubit gate,
\begin{equation}
{\mathcal{E}_{{\text{bf}}}}\left( \rho  \right) = \left( {1 - \varepsilon /2} \right)\rho  + \left( {\varepsilon /2} \right)X\rho X \ ,
\end{equation}
and (ii) a depolarizing channel, 
\begin{equation}\label{depolar_channel}
{\mathcal{E}_{{\text{dp}}}}\left( \rho  \right) = \left( {1 - \varepsilon } \right)\rho  + \varepsilon \ I/2 \ ,
\end{equation}
for each of the system qubits just before a final quantum measurement.

In fact, applying the bit-flip errors for each single-qubit gate in the ensemble is equivalent to applying a depolarizing channel before the measurement in the corresponding IQP circuit. To justify this point, let us look at Eq.~(\ref{MBQC_0J1X}) again. If a depolarizing channel is applied to the ancilla qubit before measurement, the final state after measurement is of the form (apart from a normalization factor):
\begin{equation}\label{finstateafmepi0}
\left[ {\left( {1 - \frac{\varepsilon }{2}} \right){\rho _0} + \frac{\varepsilon }{2}{\rho _1}} \right]{\Pi _0} + \left[ {\left( {1 - \frac{\varepsilon }{2}} \right){\rho _1} + \frac{\varepsilon }{2}{\rho _0}} \right]{\Pi _1} \ ,
\end{equation}
where ${\Pi _0} \equiv \left| 0 \right\rangle \left\langle 0 \right|$ and ${\Pi _1} \equiv \left| 1 \right\rangle \left\langle 1 \right|$ are projectors, ${\rho _0} = J\left( \alpha  \right)\left| \psi_\text{ini}  \right\rangle \left\langle \psi_\text{ini}  \right|J{\left( \alpha  \right)^\dag }$, and ${\rho _1} = X{\rho _0}X$. Alternatively, we can also express it as,
\begin{equation}
\mathcal{E}_{{\text{bf}}}({\rho _{{{0}}}}) \otimes {\Pi _0} + \mathcal{E}_{{\text{bf}}}({\rho _{{1}}}) \otimes {\Pi _1} \ .
\end{equation}

Therefore, we have mapped the ensemble of noisy problem of chaotic quantum circuits to the problem of an IQP circuit associated with depolarizing noise applied to each qubit before measurement. We denote the corresponding probability of obtaining the string $x$ by
\begin{equation}
{P_{{\text{exp}}}}\left( {x|{U_{x'}}} \right) \equiv \left\langle x \right|{\mathcal{E}_{x'}}\left( {{\rho _{{\text{ini}}}}} \right)\left| x \right\rangle  \ ,
\end{equation}
where ${\mathcal{E}_{x'}}$ denotes the channel describing the noisy quantum circuit labeled by $x'$. See appendix for an example. As a result, the probability $P_{{\text{IQP}}}^\varepsilon \left( {x,x'} \right)$ of getting the strings $x$ and $x'$ on a noisy IQP circuit is given by, 
\begin{equation}\label{noisyIQPdefP2m}
P_{{\text{IQP}}}^\varepsilon \left( {x,x'} \right) = \frac{1}{{{2^m}}}{P_{{\text{exp}}}}\left( {x|{U_{x'}}} \right) \ ,
\end{equation}
which is reduced to Eq.~(\ref{PIQP2mxuin}) when $\varepsilon  = 0$, i.e., the noise-free limit.

\section{--- \ Part IV: Fourier Analysis \ ---}

{\bf Fourier representation.} We shall discuss how to evaluate the probabilities, 
\begin{equation}
{P_{{\text{IQP}}}}(x,x') \equiv {\left| {\left\langle {x,x'} \right|{U_\text{IQP}}\left| {{0^{n + m}}} \right\rangle } \right|^2} \ ,
\end{equation}
of an IQP circuit classically. Here $x\in\{0,1\}^n$ is an $n$-bit string associated with the system qubits, and $x'\in\{0,1\}^m$ is for the $m$ ancilla qubits. For this purpose, we need to apply the Fourier analysis to the function $f(x,x')$ in the IQP circuit. 

Let us define 
\begin{equation}
{\chi _{s,s'}}\left( {x,x'} \right) \equiv {\left( { - 1} \right)^{s s' \cdot x x'}} \ .
\end{equation}
Any function $f(x,x')$ can be expanded by $\chi _{s,s'}$ by $f\left( {x,x'} \right) = \sum\nolimits_{s,s'} {\hat f\left( {s,s'} \right)} {\chi _{s,s'}}\left( {x,x'} \right)$, where 
\begin{equation}\label{four_co_fxx}
\hat f\left( {s,s'} \right) \equiv \frac{1}{{{2^{n + m}}}}\sum\limits_{x,x'} {f\left( {x,x'} \right){{\left( { - 1} \right)}^{s s' \cdot x x'}}} 
\end{equation}
is called the Fourier coefficient of $f(x,x')$. As a special case, 
\begin{equation}
\hat f\left( {0,0} \right) = \sum\nolimits_{x,x'} {f\left( {x,x'} \right)} /{2^{n + m}} 
\end{equation}
is simply the average value of $f(x,x')$. Comparing Eq.~(\ref{PIQP_fsum}) and (\ref{four_co_fxx}), we may also interpret the probability as given by the absolute square of the Fourier coefficient, i.e., 
\begin{equation}\label{PIQP_expform}
{P_{{\text{IQP}}}}\left( {x,x'} \right) = {| {\hat f\left( {x,x'} \right)} |^2} \ .
\end{equation}

{\bf Classical approximation of Fourier coefficients.} In the same way, the probabilities of the IQP circuit can also be expanded by its Fourier coefficients,
\begin{equation}
{{\hat P}_\text{IQP}}\left( {s,s'} \right) = \frac{1}{{{2^{n + m}}}}\sum\limits_{x,x'} {{P_\text{IQP}}\left( {x,x'} \right){{\left( { - 1} \right)}^{s s' \cdot x x'}}} \ .
\end{equation}
Now, from Eq.~(\ref{PIQP_expform}), we can choose to expand only one of the Fourier coefficients for ${P_{{\text{IQP}}}}\left( {x,x'} \right)$, i.e., ${P_{{\text{IQP}}}}\left( {x,x'} \right) = {2^{ - (n + m)}}\sum\nolimits_{yy'} {{f^*}\left( {y,y'} \right)\hat f\left( {x,x'} \right)} {\left( { - 1} \right)^{xx' \cdot yy'}}$, which gives 
\begin{equation}\label{PIQPffsum}
{{\hat P}_{{\text{IQP}}}}\left( {s,s'} \right) = \frac{1}{{{2^{2(n + m)}}}}\sum\limits_{y,y'}  \ {f_{y,y'}^*{f_{y + s,y' + s'}}}  \ ,
\end{equation}
where ${f_{x,x'}} \equiv f\left( {x,x'} \right)$.
Consequently, one can approximate the value of the Fourier coefficient ${{\hat P}_{{\text{IQP}}}}\left( {s,s'} \right)$ by uniformly sampling the product of the functions ${f_{y,y'}^*{f_{y + s,y' + s'}}}$ (and taking the real part at the end). 

From the standard Chernoff bound, for any $\eta>0$, if we take an average value from 
\begin{equation}\label{Trun_inde_trials}
T_\text{run}=O(1/\eta^2) \ ,
\end{equation}
independent trials, the approximating value, denoted by ${{\hat Q}_{{\text{cl}}}}\left( {s,s'} \right)$, is exponentially accurate with a high probability, i.e., 
\begin{equation}\label{exp_acc_hi_pro}
| {{{\hat Q}_{{\text{cl}}}}\left( {s,s'} \right) - {{\hat P}_{{\text{IQP}}}}\left( {s,s'} \right)} | \leqslant \eta \ {2^{ - \left( {n + m} \right)}} \ .
\end{equation} 

{\bf Effect of noise on Fourier coefficients.} In the following, we shall show that the effect of applying the depolarizing channel in Eq.~(\ref{depolar_channel}) on all qubits in the IQP circuit before measurement is to change each of the Fourier coefficient by some factor, i.e., 
\begin{equation}\label{FCPIQP2noisess}
{{\hat P}_{{\text{IQP}}}}\left( {s,s'} \right) \to  {\hat P_{IQP}^\varepsilon ( {s,s'} )} \equiv  {\left( {1 - \varepsilon } \right)^{| {ss'}|}} \ {{\hat P}_{{\text{IQP}}}}( {s,s'}) \ ,
\end{equation} 
where $| {ss'}|$ is the Hamming weight of the string $ss'$, i.e., the number of $1$'s. This result was stated without proof in Ref.~\cite{Bremner2016}; the following discussion provides a physical picture of this result and can be potentially extended for a generalization for further applications. As a result, the probability distribution of the IQP circuit under noise is given by,
\begin{equation}
P_{{\text{IQP}}}^\varepsilon \left( {x,x'} \right) = \sum\limits_{s,s'} {{{\left( {1 - \varepsilon } \right)}^{|ss'|}}\hat P\left( {s,s'} \right)} {\left( { - 1} \right)^{ss' \cdot xx'}} \ .
\end{equation}

First, for any given density matrix $\rho$, the operation of measurement $\mathcal{M}$ and depolarizing channel ${\mathcal{E}_{{\text{dp}}}}$ commute, i.e.,
\begin{equation}
\mathcal{M}\left( {{\mathcal{E}_{{\text{dp}}}}\left( \rho  \right)} \right) = {\mathcal{E}_{{\text{dp}}}}\left( {\mathcal{M}\left( \rho  \right)} \right) = \left( {1 - \varepsilon } \right)\mathcal{M}\left( \rho  \right) + \varepsilon I/2 \ .
\end{equation} 
Let us now consider a general $n$-qubit quantum state $\rho$ after a quantum measurement given by the following form, 
\begin{equation}
\mathcal{M}\left( \rho  \right) = \sum\limits_{x \in {{\left\{ {0,1} \right\}}^n}} {p\left( x \right)} \left| x \right\rangle \left\langle x \right| \ ,
\end{equation}
where $p\left( x \right) = \left\langle x \right|\mathcal{M}\left( \rho  \right)\left| x \right\rangle $. The same quantity can be written as,  
\begin{equation}\label{Mrho_PI_sqa}
\mathcal{M}\left( \rho  \right) = \sum\limits_{x \in {{\left\{ {0,1} \right\}}^n}} {p\left( x \right)} \ \Pi_x \ ,
\end{equation}
where $\Pi_x \equiv {\Pi _{{x_1}}}{\Pi _{{x_2}}} \cdots {\Pi _{{x_n}}} $, and ${\Pi _{{x_k}}} \equiv \left| {{x_k}} \right\rangle \left\langle {{x_k}} \right|$ for $x \in \{ 0 ,1 \}$ is a projector for the $k$-th qubit, a notation already introduced in Eq.~(\ref{finstateafmepi0}).

Next, we consider the expansion in Eq.~(\ref{Mrho_PI_sqa}) as a vector space, and introduce a Hadamard matrix $\mathcal{H}$ in the same space, i.e., for $k\in \{ 0,1\}$, $\mathcal{H} \ {\Pi _k} = ( {{\Pi _0} + {{\left( { - 1} \right)}^k} \ {\Pi _1}})/\sqrt 2$.
Note that we have the relationship,
\begin{equation}
{\mathcal{H}^{ \otimes n}} \ {\Pi _x} = \frac{1}{{\sqrt {{2^n}} }}\sum\limits_{s \in {{\left\{ {0,1} \right\}}^n}} {{{\left( { - 1} \right)}^{s \cdot x}}} \ {\Pi _s} \ ,
\end{equation}
which is similar to the standard Hadamard matrix in normal quantum circuits (see Eq.~(\ref{Hn_atacb})).

Furthermore, it is also true that, ${\mathcal{H}^2} = \mathcal{I}$ (identity). Consequently, from the fact that, $\mathcal{M}\left( \rho  \right) = {\mathcal{H}^{ \otimes n}} \ {\mathcal{H}^{ \otimes n}}\mathcal{M}\left( \rho  \right)$, we have
\begin{equation}
\mathcal{M}\left( \rho  \right) = \sqrt {{2^n}} \sum\limits_{s \in {{\left\{ {0,1} \right\}}^n}} {\hat P\left( s \right)} \ {\mathcal{H}^{ \otimes n}}{\Pi _s} \ ,
\end{equation}
where ${\mathcal{H}^{ \otimes n}}{\Pi _s} = \mathcal{H}{\Pi _{{s_1}}} \otimes \mathcal{H}{\Pi _{{s_2}}}... \otimes \mathcal{H}{\Pi _{{s_n}}}$.

\begin{proof}
Let us consider applying the depolarizing channel to one of the qubits, ${\mathcal{E}_{{\text{dp}}}}\left( {\mathcal{M}\left( \rho  \right)} \right)$. One can readily show that ${\mathcal{E}_{{\text{dp}}}}\left( {\mathcal{H}{\Pi _0}} \right) = \mathcal{H}{\Pi _0}$ and ${\mathcal{E}_{{\text{dp}}}}\left( {\mathcal{H}{\Pi _1}} \right) = \left( {1 - \varepsilon } \right)\mathcal{H}{\Pi _1}$. Therefore, for each string $s=s_1s_2 \cdots s_n$, we obtain a factor of ${\left( {1 - \varepsilon } \right)^{{s_1}}}{\left( {1 - \varepsilon } \right)^{{s_2}}} \cdots {\left( {1 - \varepsilon } \right)^{{s_n}}} \equiv {\left( {1 - \varepsilon } \right)^{|s|}}$ for each Fourier coefficient $\hat P\left( s \right)$, which completes the proof.
\end{proof}

\section{--- \ Part V: Classical approximations \ ---}

{\bf Approximating the noisy circuits.} We are now ready to approximate the noisy quantum circuits based on the family of chaotic quantum circuits encoded in the IQP circuit 
(see Eq.~(\ref{encode_IQP_QC})), in terms of the $l_1$ norm. In Eq.~(\ref{exp_acc_hi_pro}), we have seen that one can approximate each Fourier component ${{\hat P}_{{\text{IQP}}}}\left( {s,s'} \right)$ of the IQP circuit with a high probability. The question is how well can we approximate the probabilities of individual quantum circuit ${P_{{\text{qc}}}}\left( {x|{U_{x'}}} \right)$ or the experimental implementation ${P_{{\text{exp}}}}\left( {x|{U_{x'}}} \right)$? The starting point is to calculate the quantities classically, 
\begin{equation}\label{PclxxFexp_Q}
{P_{{\text{cl}}}}\left( {x,x'} \right) \equiv \sum\limits_{|ss'| \leqslant L} {{{\hat Q}^{\varepsilon}_{{\text{cl}}}}\left( {s,s'} \right)} {\left( { - 1} \right)^{ss' \cdot xx'}} \ ,
\end{equation}
which contains Fourier coefficients,
\begin{equation}\label{Qep_1messQ}
\hat Q_{{\text{cl}}}^\varepsilon \left( {s,s'} \right) \equiv {\left( {1 - \varepsilon } \right)^{|ss'|}} \ {{\hat Q}_{{\text{cl}}}}\left( {s,s'} \right) \ ,
\end{equation}
that is associated with a Hamming bound $|ss'|$ less than a given constant~$L$. The number of terms in ${P_{{\text{cl}}}}\left( {x,x'} \right)$ is bounded by 
\begin{equation}
\sum\limits_{k = 0}^L \binom{n+m}{k}    \leqslant {(n+m)^L} + 1 \ .
\end{equation}
Here each term ${{\hat Q}_{{\text{cl}}}}\left( {s,s'} \right)$ is obtained by the classical sampling method discussed below Eq.~(\ref{PIQPffsum}).

In the light of the relation shown in Eq.~(\ref{PIQP2mxuin}), we define 
\begin{equation}\label{itlrwdPcl}
{P_{{\text{cl}}}}\left( {x|{U_{x'}}} \right) \equiv {2^m}{P_{{\text{cl}}}}\left( {x,x'} \right) \ ,
\end{equation}
to be the classical approximation for the experimental implementation of $U_{x'}$ under our noise model. For a given quantum circuit $U_{x'}$, the deviation between the classical algorithm and experimental implementation can be quantified by the $l_1$ norm, i.e., ${\left\| {{P_{{\text{cl}}}}\left( {x|{U_{x'}}} \right) - {P_{\exp }}\left( {x|{U_{x'}}} \right)} \right\|_1} = \sum\nolimits_{x \in {{\left\{ {0,1} \right\}}^n}} {\left| {{P_{{\text{cl}}}}\left( {x|{U_{x'}}} \right) - {P_{\exp }}\left( {x|{U_{x'}}} \right)} \right|} $. We are interested in bounding the average performance of the classical algorithm in terms of the mean value of the $l_1$ norm over all of the quantum circuits in the ensemble, 
\begin{equation}
{\Lambda _{{\text{av}}}} \equiv (1/{2^m})\sum\limits_{x' \in {{\{ 0,1\} }^m}} {{\Lambda _{x'}}} \ ,
\end{equation}
where 
\begin{equation}\label{individ_LaxpPclPexp}
{\Lambda _{x'}} \equiv {\left\| {{P_{{\text{cl}}}}\left( {x|{U_{x'}}} \right) - {P_{{\text{exp}}}}\left( {x|{U_{x'}}} \right)} \right\|_1}\ .
\end{equation}

{\bf Bounding the average of the distributions.} 
Using Eq.~(\ref{PIQP2mxuin}), the average value ${\Lambda _{{\text{av}}}}$ can be expressed as the $l_1$ norm between difference between the probabilities of the IQP circuit under noise ${P_{{\text{IQP}}}^\varepsilon \left( {x,x'} \right)}$ (see Eq.~(\ref{noisyIQPdefP2m})) and the classical algorithm ${{P_{{\text{cl}}}}\left( {x,x'} \right)}$ approximating it, i.e., 
\begin{equation}
\Lambda_\text{av} = \left\| {{P_{{\text{cl}}}}\left( {x,x'} \right) - P_{{\text{IQP}}}^\varepsilon \left( {x,x'} \right)} \right\| \ .
\end{equation}
The goal of this section is to show there exists a polynomial-time classical algorithm such that $\Lambda_\text{av}$ can be bounded by a small constant $\delta$, i.e., 
\begin{equation}
{\Lambda _{{\text{av}}}} \leqslant c \ \delta \ ,
\end{equation}
for some constant $c$.

For any vector, ${\bf x}=(x_1,x_2,..,x_n)^T$, the $l_1$ norm, ${\left\| {\mathbf{x}} \right\|_1} = \sum\nolimits_{i = 1}^n {\left| {{x_i}} \right|} $, can always be bounded by the $l_2$ norm, ${\left\| {\mathbf{x}} \right\|_2} = {({\sum\nolimits_{i = 1}^n {|{x_i}|} ^2})^{1/2}}$, i.e., ${\left\| {\mathbf{x}} \right\|_1} \leqslant \sqrt n {\left\| {\mathbf{x}} \right\|_2}$. Therefore, we can write, 
\begin{equation}\label{LavlPclIQP}
\Lambda _{{\text{av}}}^2 \leqslant {2^{n + m}}\sum\limits_{x,x'} {{{\left( {{P_{{\text{cl}}}}\left( {x,x'} \right) - P_\text{IQP}^\varepsilon \left( {x,x'} \right)} \right)}^2}} \ .
\end{equation}
The right-hand side can be replaced by the corresponding Fourier coefficients using Parseval's identity, which gives 
\begin{equation}
\Lambda _{{\text{av}}}^2 \leqslant {2^{2(n + m)}}\sum\nolimits_{s,s'} {{{( {{{\hat P}_{{\text{cl}}}}\left( {s,s'} \right) - \hat P_{IQP}^\varepsilon \left( {s,s'} \right)} )}^2}} \ .
\end{equation}
Note that there is an extra factor of $2^{n+m}$. 

Recall in Eq.~(\ref{PclxxFexp_Q}) that ${{\hat P}_{{\text{cl}}}}\left( {s,s'} \right) = \hat Q_{{\text{cl}}}^\varepsilon \left( {s,s'} \right)$ for the Hamming distance of the string $ss'$ to be less than a constant $L$; otherwise ${{\hat P}_{{\text{cl}}}}\left( {s,s'} \right) = 0$. Recall also in Eq.~(\ref{Qep_1messQ}) that there are $O(n+m)^L$ non-zero terms. Therefore, we divide the summation into two parts, i.e.,
\begin{equation}
\Lambda _{{\text{av}}}^2 \ \leqslant \ {\Omega _1} + {\Omega _2} \ ,
\end{equation}
where the first term is given by, 
\begin{equation}
{\Omega _1} \equiv {2^{2(n + m)}}\sum\limits_{|ss'| \leqslant L} {{{( {{{\hat P}_{{\text{cl}}}}\left( {s,s'} \right) - \hat P_\text{IQP}^\varepsilon \left( {s,s'} \right)} )}^2}}  \ ,
\end{equation}
and the second term can be written as,
\begin{equation}
{\Omega _2} \equiv {2^{2(n + m)}}\sum\limits_{|ss'| > L} {\hat P_\text{IQP}^\varepsilon {{\left( {s,s'} \right)}^2}} \ ,
\end{equation} 
which (using Eq.~(\ref{FCPIQP2noisess})) can be bounded by 
\begin{equation}
{\Omega _2} \leqslant {2^{2(n + m)}}{\left( {1 - \varepsilon } \right)^{2L}}\sum\limits_{s,s'} {{{\hat P}_{{\text{IQP}}}}{{\left( {s,s'} \right)}^2}} \ .
\end{equation}

Let us further investigate the two $\Omega$ terms separately; the following analysis is similar to the one performed in Ref.~\cite{Bremner2016}; we provide the details in terms of our notations. For the term $\Omega_1$, since ${\left( {1 - \varepsilon } \right)^{|ss'|}} \leqslant 1$, we have 
\begin{equation}
| {{{\hat P}_{{\text{cl}}}}\left( {s,s'} \right) - \hat P_{{\text{IQP}}}^\varepsilon \left( {s,s'} \right)} | \leqslant | {{{\hat Q}_{{\text{cl}}}}\left( {s,s'} \right) - {{\hat P}_{{\text{IQP}}}}\left( {s,s'} \right)} | \ .
\end{equation}
Together with  Eq.~(\ref{exp_acc_hi_pro}), we conclude that, 
\begin{equation}
{\Omega _1} \leq  {{\eta ^2}{((n+m)^L+1)}} \ .
\end{equation}
In order to have it bounded by a small constant, e.g., ${\Omega _1} \leq  {{\delta ^2}}$, we need to make, 
\begin{equation}
\eta  = O(\delta /{(n + m)^{L/2}}) \ ,
\end{equation}
which requires the runtime of the Monte Carlo algorithm to scales as (see Eq.~(\ref{Trun_inde_trials}))
\begin{equation}
T_\text{run} = O( {{{(n + m)}^L}/{\delta ^2}} ) \ .
\end{equation}
The value of $L$ is determined by the second term.

For the second term, we can now go back to the standard basis, i.e., 
\begin{equation}
{\Omega _2} \leqslant {2^{n + m}}{\left( {1 - \varepsilon } \right)^{2L}}\sum\limits_{x,x'} {{P_{{\text{IQP}}}}{{\left( {x,x'} \right)}^2}} \ .
\end{equation}
Using Eq.~(\ref{PIQP2mxuin}), we have 
\begin{equation}
{\Omega _2} \leqslant {\left( {1 - \varepsilon } \right)^{2L}}{2^{n - m}}\sum\limits_{x,x'} {{P_{{\text{qc}}}}{{\left( {x|{U_{x'}}} \right)}^2}} \ ,
\end{equation}
 which implies that,  
\begin{equation}
{\Omega _2} \leqslant {\left( {1 - \varepsilon } \right)^{2L}}{2^{n - m}}\sum\limits_{x'} {{R_{x'}}}  \leqslant \alpha {\left( {1 - \varepsilon } \right)^{2L}} \ ,
\end{equation}
with the use of the assumption made in Eq.~(\ref{assum_Q2n}). Unless $\alpha$ is of order $\delta^2$ or smaller, we need to make $L$ to be sufficiently large, so that $\alpha {\left( {1 - \varepsilon } \right)^{2L}} \leqslant \alpha {e^{ - 2 \varepsilon L}} = {\delta ^2}$, and hence 
\begin{equation}
L = O(\log (\alpha/{\delta ^2})/\varepsilon ) \ .
\end{equation}
As a result, the average $l_1$ norm ${\Lambda _{{\text{av}}}}$ can be bounded by the classical polynomial-time algorithm to a small constant, i.e., 
\begin{equation}\label{PclPIQPcd2}
\sum\limits_{x,x'} {{{\left( {{P_{{\text{cl}}}}\left( {x,x'} \right) - P_{{\text{IQP}}}^\varepsilon \left( {x,x'} \right)} \right)}^2}}  \leqslant \frac{{c^2 \delta^2 }}{{{2^{n + m}}}} \ ,
\end{equation}
and hence 
${\Lambda _{{\text{av}}}} \leqslant c \  \delta$ from Eq.~(\ref{LavlPclIQP}).

{\bf Bounding the norm for each circuit.} We have shown that the average value of the $l_1$ norms, $\Lambda_\text{av}$, can be bounded by a small constant by a polynomial-time classical algorithm. Next, we can further ask the following question: if we randomly pick one of the quantum circuits, $U_{x'}$, what is the probability that the classical algorithm fails to maintain an  $l_1$ norm close to the average value, i.e., within a constant multiple of~$\delta$; the answer to this question depends on the variance of the distribution of $\Lambda_{x'}$ defined in Eq.~(\ref{individ_LaxpPclPexp}).

Recall that for a random variable $X$, the Chebyshev inequality states that 
\begin{equation}
\Pr \left( {\left| {X - \mu } \right| \geqslant \lambda } \right) \leqslant {\text{Var}}\left( X \right)/{\lambda ^2} \ ,
\end{equation}
where $\mu  \equiv \sum\nolimits_a a \Pr \left( {X = a} \right)$ is the mean value, and ${\text{Var}}\left( X \right) = \langle {{{\left| X \right|}^2}} \rangle  - {\left| \mu  \right|^2}$ is the variance. For our case, we set $X=\Lambda_{x'}$, and hence $\mu  = {\Lambda _{{\text{av}}}}$. 

Furthermore, we have 
\begin{equation}
{\text{Var}}\left( X \right) \leqslant \langle {{X^2}} \rangle  = {2^{ - m}}\sum\limits_{x'} {\Lambda _{x'}^2} \ ,
\end{equation}
 where 
\begin{equation}
\Lambda _{x'}^2 \leqslant {2^n}\sum\limits_x {{{({P_{{\text{cl}}}}\left( {x|{U_{x'}}} \right) - {P_{\exp }}\left( {x|{U_{x'}}} \right))}^2}} \ ,
\end{equation}
using again ${\left\| {\mathbf{x}} \right\|_1} \leqslant \sqrt n {\left\| {\mathbf{x}} \right\|_2}$. From Eq.~(\ref{itlrwdPcl}) and (\ref{noisyIQPdefP2m}), we have 
\begin{equation}
\langle {{X^2}} \rangle  \leqslant {2^{n + m}}\sum\limits_{x,x'} {{{({P_{{\text{cl}}}}\left( {x,x'} \right) - P_{{\text{IQP}}}^\varepsilon \left( {x,x'} \right))}^2}} \ ,
\end{equation}
 which is exactly the right-hand side of Eq.~(\ref{LavlPclIQP}). Therefore, we have 
\begin{equation}
\langle X^2 \rangle \leq 2\delta^2 \ .
\end{equation}
Now, if we set $\lambda = k \, \delta$, then the probability for the $l_1$ norm $\Lambda_{x'}$ to be deviated from the mean value by an amount of $k \delta$ is less than $2/k^2$, i.e., 
\begin{equation}\label{PrLxpmLavk2}
\Pr \left( {\left| {{\Lambda _{x'}} - {\Lambda _{{\text{av}}}}} \right| \geqslant k\delta } \right) \leqslant 2/{k^2} \ ,
\end{equation}
which can become very small, e.g. by making $k=10$.

{\bf Classical sampling algorithm.} So far, we have explained how to obtain an numerical approximation of the probability ${P_{{\text{exp}}}}\left( {x|{U_{x'}}} \right)$ for a given $x$ and $U_{x'}$. In order to produce a ``distribution" in practice, we need a sampling algorithm that can be justified to be an accurate approximation of the noisy quantum circuits. The sampling algorithm we shall be discussing is a modified version of the one described in Ref.~\cite{Bremner2016}. However, we shall present the algorithm in an alternative, and from our point of view, more direct approach.

Our goal is to show how to produce classically the distribution described by any $U_{x'}$ in the quantum ensemble under noise, i.e., bit strings following closely the distribution described by~${P_{{\text{exp}}}}\left( {x|{U_{x'}}} \right)$. The challenge is that there is no guarantee that the classical approximation of the probabilities are necessarily positive numbers, through the truncation of the Fourier coefficients~Eq.~(\ref{PclxxFexp_Q}). Fortunately, as we shall show that such a difficulty can be overcome.  

The classical sampling algorithm proposed in Ref.~\cite{Bremner2016} is quite simple; it is a random walk, involving an application of two repeating steps. Each time the walker has to decide randomly to take the next step to be `0' or `1'. The position of the walk is described by a binary string of variable length. To describe the algorithm in detail, suppose after $k$ steps, the walker stopped at a position labeled by $z_k \equiv x_1 x_2 \cdots x_k$. 

{\bf Step 1:} calculate the pseudo-probability (omitting the label of $x'$ for simplicity), 
\begin{equation}
p\left( {{z_k}} \right) \equiv \sum\limits_{x,{z_k}} {{P_{{\text{cl}}}}\left( {x} \right)} \ ,
\end{equation}
for any given partial string, which is the sum of all calculated approximation of the probabilities subject to the constraint that the first $k$ bits are fixed to be $z_k$. Note that when $k=n$, 
\begin{equation}
p\left( {{z_n}} \right) = {P_{{\text{cl}}}}\left( {{z_n}} \right) = {P_{{\text{cl}}}}\left( {{x_1}{x_2} \cdots {x_n}} \right) \ .
\end{equation}

Similarly, the values of $p\left( {{z_k}0} \right)$ and $p\left( {{z_k}1} \right)$ are also needed, where 
\begin{equation}\label{pz_pz0_pz1}
p\left( {{z_k}} \right) = p\left( {{z_k}0} \right) + p\left( {{z_k}1} \right) \ ,
\end{equation} 
by definition. These values can be determined by evaluating the Fourier coefficients (see Ref.~\cite{Shepherd2010,Bremner2016}). Note that the values of $p\left( {{z_k}0} \right)$ and $p\left( {{z_k}1} \right)$ may be negative, but not both, i.e., $p\left( {{z_k}} \right)$ is necessarily positive, from the construction of the next step. 

{\bf Step 2:} check if [case 1:] both $p(z_k0)$ and $p(z_k1)$ are positive, or [case 2:] one of them is negative. If both values are positive, then the walker takes `0' or `1' based on the following probabilities, 
\begin{equation}
\Pr \left( 0 \right) = \frac{{p\left( {{z_k}0} \right)}}{{p\left( {{z_k}} \right)}},\quad \Pr \left( 1 \right) = \frac{{p\left( {{z_k}1} \right)}}{{p\left( {{z_k}} \right)}} \ .
\end{equation}

However, if one of them is negative, e.g., $p\left( {{z_k}0} \right) > 0$ and $p\left( {{z_k}1} \right) < 0$, then the walker chooses the positive side with deterministically, i.e., 
\begin{equation}
\Pr \left( 0 \right) = 1, \quad \Pr \left( 1 \right) = 0 \ .
\end{equation}

{\bf Performance and the sampling algorithm.} To analyze the performance of the sampling algorithm, we suppose, after $k-1$ steps, the distribution generated by the algorithm is given by,
\begin{equation}
{\text{Alg}}_{k - 1} = {a_0}{\Pi _0} + {a_1}{\Pi _1} + ... + {a_m}{\Pi _m} \ ,
\end{equation}
where $\Pi_i$ labels one of the projectors, and for all $a_i > 0$, we have 
\begin{equation}\label{a0a1dddam_fawh}
{a_0} + {a_1} + ... + {a_m} = 1 \ . 
\end{equation}
We shall compare the distribution with a mathematical vector calculated classically, 
\begin{equation}\label{Pmath_vec_cal_cl}
P_{{\text{math}}}^{\left( k \right)} \equiv \left( {{p_0}{\Pi _0} + {p_1}{\Pi _1} + ... + {p_{{2^k} - 1}}{\Pi _{{2^k} - 1}}} \right)/S \ ,
\end{equation}
where value of the $p_i$ is taken to be one of the $p(z_k)$ in Eq.~(\ref{pz_pz0_pz1}), and 
\begin{equation}
S = {p_0} + {p_1} + ... + {p_{{2^k} - 1}} 
\end{equation}
is a normalization factor. Note that the value of $S$ is in fact independent of $k$, as $S = \sum\nolimits_x {{P_{{\text{cl}}}}\left( x \right)} $.

Furthermore, from the discussion around Eq.~(\ref{PrLxpmLavk2}), there is a high probability to find a quantum circuit $U_{x'}$, such that the classically-calculated norm is bounded by $k \delta$, i.e., 
\begin{equation}\label{PexpPclkd_cnibb}
\sum\limits_x {\left| {{P_{\exp }}\left( x \right) - {P_{{\text{cl}}}}\left( x \right)} \right|}  \leqslant k \delta \ .
\end{equation}
The left-hand side is larger than the following: 
\begin{equation}
|\sum\nolimits_x ({{P_{\exp }}\left( x \right) - {P_{{\text{cl}}}}\left( x \right)}) | = \left| {1 - S} \right| \ ,
\end{equation}
which means that $\left| {1 - S} \right| \leqslant k\delta $, or equivalently, 
\begin{equation}\label{1kS1k_wmt}
1 + k\delta  \geqslant S \geqslant 1 - k\delta \ .
\end{equation}
Practically, we would need to choose $k \delta \ll 1$.

Now, we shall show that for each term, 
\begin{equation}\label{ax_pxS_4et}
{a_x} \leq {p_x}/S  \ ,
\end{equation}
which means that the algorithm cannot generate a probability larger than the calculated value. The proof can be achieved by induction.
\begin{proof}
Suppose, at some point, it is true that ${a_x} = {g_x}{p_x}/S$ for some $g_x$, where $0 < {g_x} \leqslant 1$. In the next step, we consider the values of $p_{x0}$ and $p_{x1}$, which are the calculated values of the next step, i.e., $p\left( {{z_k}0} \right)$ and $p\left( {{z_k}1} \right)$ in Eq.~(\ref{pz_pz0_pz1}). According to the mechanism of the sampling algorithm, if both terms are positive, i.e., $p\left( {{z_k}0} \right) > 0$ and $p\left( {{z_k}1} \right) > 0$. Then we set ${c_x}{\Pi _x} \to {g_x}\left( {{p_{x0}}/S} \right){\Pi _{x0}} + {g_x}\left( {{p_{x1}}/S} \right){\Pi _{x1}}$. Compared with the vector in Eq.~(\ref{Pmath_vec_cal_cl}), we have $\left( {{p_{x0}}/S} \right) - {g_x}\left( {{p_{x0}}/S} \right) \geqslant 0$ and $\left( {{p_{x1}}/S} \right) - {g_x}\left( {{p_{x1}}/S} \right) \geqslant 0$. 

Now, if one of them is negative (recall that it is impossible to have both terms negative), e.g., ${p_{x0}} > 0$ and ${p_{x1}} < 0$. We then have ${c_x}{\Pi _x} \to {g_x}\left( {{p_x}/S} \right){\Pi _{x0}} = {g_x}\left( {{p_x}/{p_{x0}}} \right)\left( {{p_{x0}}/S} \right){\Pi _{x0}}$. Note that ${p_x} = {p_{x0}} + {p_{x1}} < {p_{x0}}$. Therefore, the value ${g_x}\left( {{p_x}/{p_{x0}}} \right)\left( {{p_{x0}}/S} \right)$ is also smaller than that $\left( {{p_{x0}}/S} \right)$ of $P_{{\text{math}}}^{\left( k \right)}$.
\end{proof}

At the end, i.e., when $k=n$, we have $p_x = P_\text{cl}(x)$. From the result of Eq.~(\ref{ax_pxS_4et}), the $l_1$ norm between the sampling algorithm and the calculation can be written as, 
\begin{equation}
{\left\| {{P_{{\text{math}}}} - {\text{Alg}}} \right\|_1} = \sum\limits_{x,{p_x} > 0} {\left( {\frac{{{p_x}}}{S} - {a_x}} \right)}  + \sum\limits_{x,{p_x} < 0} {\frac{{\left| {{p_x}} \right|}}{S}} \ ,
\end{equation}
where the first summation contains all positive terms. The second summation contains the negative terms. Now, we can use Eq.~(\ref{a0a1dddam_fawh}) to make 
\begin{equation}
\sum\limits_{x,{p_x} > 0} {{a_x}}  = 1 = \frac{S}{S} = \frac{1}{S}\sum\limits_x {{p_x}} \ .
\end{equation}
Note that the summation on the right includes all possible values of $x$. Consequently, with a high probability, we have 
\begin{equation}
{\left\| {{P_{{\text{math}}}} - {\text{Alg}}} \right\|_1} = \frac{2}{S}\sum\limits_{x,{p_x} < 0} {\left| {{p_x}} \right|}  \leqslant \frac{{2  \, k \, \delta }}{{1 - k \, \delta }} \ ,
\end{equation} 
where we used Eq.~(\ref{1kS1k_wmt}), and Eq.~(\ref{PexpPclkd_cnibb}) in the following way: 
\begin{equation}
\sum\limits_{x,{p_x} < 0} {\left| {{p_x}} \right|}  \leqslant \sum\limits_{x,{p_x} < 0} {\left| {{P_{\exp }}\left( x \right) - {P_{{\text{cl}}}}\left( x \right)} \right|}  \leqslant k\delta \ .
\end{equation}

Finally, we can ask how good is the sampling algorithm compared with the experimental implementation. We first consider the $l_1$ norm, i.e., 
\begin{equation}
{\left\| {{\text{Alg}} - {P_{{\text{exp}}}}} \right\|_1} \leqslant {\left\| {{P_{{\text{math}}}} - {\text{Alg}}} \right\|_1} + {\left\| {{P_{{\text{math}}}} - {P_{{\text{exp}}}}} \right\|_1} \ ,
\end{equation}
using the triangle inequality, where 
\begin{equation}
{\left\| {{P_{{\text{math}}}} - {P_{{\text{exp}}}}} \right\|_1} = \sum\limits_x {\left| {{P_{{\text{cl}}}}\left( x \right)/S - {P_{\exp }}\left( x \right)} \right|} \ .
\end{equation}
Applying again the triangle inequality, we have ${\left\| {{P_{{\text{math}}}} - {P_{{\text{exp}}}}} \right\|_1} \leqslant \frac{1}{S}{\left\| {{P_{{\text{cl}}}} - {P_{{\text{exp}}}}} \right\|_1} + \left| {\frac{1}{S} - 1} \right|{\left\| {{P_{{\text{exp}}}}} \right\|_1}$, which implies that 
\begin{equation}
{\left\| {{P_{{\text{math}}}} - {P_{{\text{exp}}}}} \right\|_1} \leqslant \frac{{k\delta }}{S} + \frac{{\left| {1 - S} \right|}}{S} \leqslant \frac{{2k\delta }}{{1 - k\delta }} \ ,
\end{equation}
from Eq.~(\ref{PexpPclkd_cnibb}) and Eq.~(\ref{1kS1k_wmt}). Consequently, we have
\begin{equation}
{\left\| {{\text{Alg}} - {P_{{\text{exp}}}}} \right\|_1} \leqslant \frac{{4 \, k \, \delta }}{{1 - k \, \delta }} \ .
\end{equation}

{\bf Bounding the cross entropy.} Now, we are ready to consider the cross entropy. In Eq.~(\ref{DSScScw}), we argued that the main quantity of interest is the difference in cross entropy, $\Delta_{S} \equiv \left| {{S_c}\left( {{P_{\exp }},{P_{{\text{qc}}}}} \right) - {S_c}\left( {{P_{{\text{cl}}}},{P_{{\text{qc}}}}} \right)} \right|$, which is bounded by the following:
\begin{equation}
{\Delta _S} \leqslant \sum\limits_{x \in {{\left\{ {0,1} \right\}}^n}} {\left| {{P_{\exp }}\left( x \right) - {P_{{\text{cl}}}}\left( x \right)} \right| \cdot \left| {\log {P_{{\text{qc}}}}\left( x \right)} \right|} \ ,
\end{equation}
which can be extended to any circuit $U_{x'}$ in the ensemble, e.g., ${P_{\exp }}\left( x \right) \to {P_{\exp }}\left( {x|{U_{x'}}} \right)$. Let us now consider an ensemble of this quantity, 
\begin{equation}
{E_\Delta } \equiv \frac{1}{{{2^m}}}\sum\limits_{x'} {{\Delta _{S,x'}}} \ ,
\end{equation}
where ${\Delta _{S,x'}}$ denotes the ${\Delta _{S}}$ associated with one of the quantum circuits $U_{x'}$ in the ensemble. 

Using Eq.~(\ref{noisyIQPdefP2m}) and Eq.~(\ref{itlrwdPcl}), the average value is bounded by the following: 
\begin{equation}
{E_\Delta }\leqslant \sum\limits_{x,x'} {A\left( {x,x'} \right)B\left( {x,x'} \right)} \ ,
\end{equation}
where $A\left( {x,x'} \right) \equiv \left| {P_{{\text{IQP}}}^\varepsilon \left( {x,x'} \right) - {P_{{\text{cl}}}}\left( {x,x'} \right)} \right|$ and $B\left( {x,x'} \right) \equiv \left| {\log {P_{{\text{qc}}}}\left( {x|{U_{x'}}} \right)} \right|$.
Applying the Cauchy-Schwarz inequality to the upper bound of ${E_\Delta }$, we have $E_\Delta ^2 \leqslant \sum\nolimits_{x,x'} {A{{\left( {x,x'} \right)}^2}\sum\nolimits_{x,x'} {B{{\left( {x,x'} \right)}^2}} }$. According to Eq.~(\ref{PclPIQPcd2}), 
\begin{equation}
\sum\limits_{x,x'} {A{{\left( {x,x'} \right)}^2}} \leqslant \frac{{{c^2}{\delta ^2}}}{{{2^{n + m}}}} \ .
\end{equation} 

On the other hand, assuming the Porter-Thomas distribution for the quantum circuits, we have (see appendix)
\begin{equation}
\sum\limits_{x,x'} {B{{\left( {x,x'} \right)}^2}}  = {2^{n + m}}\left[ {{{\left( {n\log 2 + \gamma } \right)}^2} + {\pi ^2}/6} \right] \ .
\end{equation}
As a result, we have
\begin{equation}
{E_\Delta } = O\left( {\delta \sqrt {{{\left( {n\log 2 + \gamma } \right)}^2} + {\pi ^2}/6} } \right) \ .
\end{equation}
Therefore, the average value can be bounded as above. To keep it a constant, it is sufficient to require $\delta$ to scale as $1/n$. This completes our analysis.

\section{Conclusion}
In this work, we consider the problem of demonstrating the quantum supremacy for a  family of chaotic quantum circuits, subject to noise. For chaotic circuits obeying the Porter-Thomas distribution, we found that there exist a polynomial-time classical algorithm that can simulate the average distribution within a constant $l_1$ norm. Furthermore, the classical algorithm becomes quasi-polynomial if we further require it to bound the difference in cross entropy within a small constant.

Going back to the original question, can chaotic quantum circuits maintain quantum supremacy under noise? Our results suggest that it really depends on our knowledge on the noise and the strength $\varepsilon$ of the noise. In the extreme case, where the quantum circuits are subject to very strong depolarizing noise, the final state becomes very close to a completely-mixed state, and therefore becomes simulable classically. Indeed, our bound is very sensitive to the value of $\varepsilon$. The runtime of the classical algorithm scales as $n^{1/\varepsilon}$; it becomes a very expensive polynomial when $\varepsilon$ approaches zero. Therefore, the answer to the question depends how much can we improve the classical algorithm, or how small the $\varepsilon$ is. The bottom line is that our results highlight the challenges  for an experimental demonstration of quantum supremacy. We believe that a better theoretical understanding of the problem is needed before one can design a ``loop-hole free" experimental demonstrating on quantum supremacy.

{\it Acknowledgments---} M.-H.Y. acknowledges the support by the National Natural Science Foundation of China under Grants No. 11405093, and Guangdong Innovative and Entrepreneurial Research Team Program (No.2016ZT06D348). X. G. acknowledges the support by the National key Research
and Development Program of China.

\bibliographystyle{apsrev4-1}
\bibliography{reference}

\newpage 

\appendix

\section{Appendix: Cross entropy}
Here we review the emergence of cross entropy. Let 
\begin{equation}
\left| \psi  \right\rangle  = U\left| {{\psi _0}} \right\rangle \ ,
\end{equation}
be the output of a given random quantum circuit, $x_j^\text{qc}$ be a bit string obtained from a quantum measurement on the final state in the computational basis $\left\{ {\left| {{x_j}} \right\rangle } \right\}$,
\begin{equation}
s_\text{qc} = \left\{ {{x_1^\text{qc}},{x_2^\text{qc}},...,{x_m^\text{qc}}} \right\}
\end{equation}
 be a sequence obtained by $m$ quantum measurements. The probability of obtaining the sequence $s_{\rm qc}$ is given by the product, 
\begin{equation}
\Pr \left( {{s_{{\text{qc}}}}} \right) = \prod\limits_{i = 1}^m {{P_{{\text{qc}}}}\left( {x_i^{{\text{qc}}}} \right)} \ ,
\end{equation}
where ${{P_{{\text{qc}}}}\left( {x} \right)} \equiv {\left| {\left\langle x \right|\left. \psi  \right\rangle } \right|^2} $.
Let us now consider (removing the superscript for simplicity),
\begin{equation}
\log \Pr \left( {{s_{{\text{qc}}}}} \right) = m \times \frac{1}{m}\sum\limits_{{x_j} \in s_\text{qc}} {\log } \ {P_\text{qc}}\left( {{x_j}} \right) \ .
\end{equation}
If we uniformly and randomly pick the $x_j$'s, then from the central limit theorem, 
\begin{equation}\label{uniform_ran}
\log \Pr \left( {{s_{{\text{qc}}}}} \right) =  - m \ S \left( {{P_\text{qc}}} \right) + O( {{m^{1/2}}}) \ ,
\end{equation}
where
\begin{equation}
S \left( {{P_\text{qc}}} \right) \equiv  - \sum\limits_{j = 1}^N {{P_\text{qc}}\left( {{x_j}} \right)\log } \ {P_\text{qc}}\left( {{x_j}} \right) \ ,
\end{equation}
is the Shanon entropy.

Now, let ${s_\text{cl}} = \left\{ {x_1^\text{cl},x_2^\text{cl},...,x_m^\text{cl}} \right\}$ be the sequence of bit strings generated by a classical algorithm. Let us consider 
\begin{equation}
{\Pr}\left( {{s_\text{cl}}} \right) = \prod\limits_{{x_j} \in {s_\text{cl}}} {{P_\text{qc}}\left( {x_j^\text{cl}}  \right)} \ ,
\end{equation}
the joint probability the quantum circuit would produce the same sequence $s_\text{cl}$. Taking the logarithm, we have 
\begin{equation}
\log {\Pr} \left( {{s_\text{cl}}} \right) = \sum\limits_{{x_j} \in {s_\text{cl}}} {\log {P_\text{qc}}\left( {{x_j}} \right)} \ .
\end{equation}  

Following the procedure in equation~(\ref{uniform_ran}), where the central limit theorem was applied, we have
\begin{equation}
\log {\Pr} \left( {{s_\text{cl}}} \right) =  - m \ S_c \left( {{P_\text{cl}},{P_\text{qc}}} \right) + O( {{m^{1/2}}}) \ ,
\end{equation}
where 
\begin{equation}
S_c \left( {{P_\text{cl}},{P_\text{qc}}} \right) \equiv  - \sum\limits_{j = 1}^N {{P_\text{cl}(x_j)} \log {P_\text{qc}}} \left( {{x_j}} \right) \ ,
\end{equation}
is the cross entropy between the strings generated by the classical algorithm and quantum circuit.

Similarly, for a sequence of strings ${s_{\exp }}$ produced by an experiment, we have
\begin{equation}
\log {\Pr} \left( {{s_{\exp }}} \right) =  - m \ S_c \left( {{P_{\exp }},{P_\text{qc}}} \right) + O( {{m^{1/2}}}) \ .
\end{equation}

\section{Appendix: Porter-Thomas and the second moment}

Consider the second moment of the probabilities $P(x)$ in sampling the bit-strings $x\in\{0,1\}^n$ of a general quantum circuit:
\begin{equation}
\sum\limits_{x \in {{\left\{ {0,1} \right\}}^n}} {P{{\left( x \right)}^2}}  = \sum\limits_{i = 1}^N {P{{\left( {{x_i}} \right)}^2}}  \equiv \sum\limits_{i = 1}^N {P_i^2} ,
\end{equation}
where we labelled the probabilities as ${P_i} \equiv P\left( {{x_i}} \right)$ for simplicity, and set $N=2^n$. 

Of course, we can also express it as follows: 
\begin{equation}\label{SP2IfP2oc}
\sum\limits_{i = 1}^N {P_i^2}  = \int_0^\infty  {f\left( P \right)} \ {P^2} \ dP \ ,
\end{equation}
where the distribution function is given by,
\begin{equation}
f\left( P \right) \equiv \sum\limits_{i = 1}^N {\delta \left( {P - {P_i}} \right)} \ .
\end{equation}

Note that the normalization of $f(P)$ has to be $N$ instead of $1$, as 
\begin{equation}
\int_0^\infty  {f\left( P \right)} dP = \int_0^\infty  {\sum\limits_{i = 1}^N {\delta \left( {P - {P_i}} \right)} dP = N} \ .
\end{equation}
Therefore, for the Porter-Thomas distribution, we have to make 
\begin{equation}\label{fPN2emNPPtd}
f\left( P \right) = {N^2}{e^{ - NP}} \ ,
\end{equation}
which gives the second moment,
\begin{equation}
\left\langle {{P^2}} \right\rangle  \equiv \int_0^\infty  {f\left( P \right)} {P^2}dP = 2/N \ .
\end{equation}
In general, the $k$-th moment of the Porter-Thomas distribution is given by, 
\begin{equation}
\left\langle {{P^k}} \right\rangle  = {N^{ - k + 1}}k! \ ,
\end{equation}
which can be obtained by a similar argument.

\section{Appendix: noise conversion for two qubits}

Let us consider an extension of the Eq.~(\ref{MBQC_0J1X}) for the case of two qubits for a further elaboration of how depolarizing noise can be converted into Pauli noise, 
\begin{equation}
\left| {{\psi _J}} \right\rangle  = \frac{1}{2}\sum\limits_{x' \in {{\left\{ {0,1} \right\}}^2}} {{U_{x'}}\left| \psi_{\rm ini}  \right\rangle \otimes \left| {x'} \right\rangle } \ ,
\end{equation}
where ${U_{00}} = J_2\left( \beta  \right)J_1\left( \alpha  \right)$, ${U_{01}} = J_2 \left( \beta  \right) X J_1\left( \alpha  \right)$, ${U_{10}} = X J_2\left( \beta  \right) J_1\left( \alpha  \right)$, and  ${U_{11}} = X J_2\left( \beta  \right)XJ_1\left( \alpha  \right)$.

Suppose a single-qubit depolarizing channel is applied to the first ancilla qubits, after the measurement, we have the resulting state,
\begin{equation}
\sum\limits_{x,y \in \left\{ {0,1} \right\}} {\left( {\left( {1 - \frac{\varepsilon }{2}} \right){\rho _{xy}} + \frac{\varepsilon }{2}{\rho _{\bar xy}}} \right) \otimes {\Pi _{xy}}}  \ ,
\end{equation}
where ${\bar x}$ represents the complement of $x$, and 
\begin{equation}
{\rho _{xy}} \equiv {U_{xy}} \ {\rho _{{\text{ini}}}} \ U_{xy}^\dag  \ .
\end{equation}
This expression can be viewed as an application of the bit-flipping channel to the first qubit, i.e., 
\begin{equation}
\sum\limits_{x,y \in \{ 0,1\} } {\left( {{\mathcal{E}_{{\text{bf}}}} \otimes I} \right)({U_{xy}}{\rho _{{\text{ini}}}}U_{xy}^\dag ) \otimes {\Pi _{xy}}} \ .
\end{equation}

Let us apply the depolarizing channel to the second ancilla qubit, giving 
\begin{equation}
\sum\limits_{x,y} {\left( {\left( {1 - \frac{\varepsilon }{2}} \right)\left( {{\mathcal{E}_{{\text{bf}}}} \otimes I} \right){\rho _{xy}} + \frac{\varepsilon }{2}\left( {{\mathcal{E}_{{\text{bf}}}} \otimes I} \right){\rho _{x\bar y}}} \right) \otimes {\Pi _{xy}}} \ .
\end{equation}
For example, let us focus on the $\Pi_{00}$ term, which is associated with the state,
\begin{equation}
p_\varepsilon ^2 \ {\rho _{00}} + {p_\varepsilon }{q_\varepsilon } \ {\rho _{10}} + {q_\varepsilon }{p_\varepsilon } \ {\rho _{01}} + q_\varepsilon ^2 \ {\rho _{11}} \ ,
\end{equation}
where  ${p_\varepsilon } \equiv 1 - \varepsilon /2$ and ${q_\varepsilon } \equiv \varepsilon /2$. 

Therefore, if we measure the ancilla qubits and obtain the outcome $00$, then it is equivalent to the case where a quantum circuit $U_{00}$ is implemented under the bit-flip noises after applying each $J$ gate. In general, given $x'$ from the ancilla qubits, the probability of getting a bit-string $x$ is given by 
\begin{equation}
{P_{{\text{exp}}}}\left( {x|{U_{x'}}} \right) \equiv \left\langle x \right|{\mathcal{E}_{x'}}\left( {{\rho _{{\text{ini}}}}} \right)\left| x \right\rangle  \ ,
\end{equation}
where ${\mathcal{E}_{x'}}$ labels the quantum channel of the noisy quantum circuit. For example, if $x'=00$, we have ${\mathcal{E}_{00}}\left( \rho  \right) = \left( {I \otimes {\mathcal{E}_{{\text{bf}}}}} \right)({J_2}\left( \beta  \right)\left( {{\mathcal{E}_{{\text{bf}}}} \otimes I} \right)({J_1}\left( \alpha  \right)\rho {J_1}{\left( \alpha  \right)^\dag }){J_2}{\left( \beta  \right)^\dag })$. Note that in the main text, the quantity ${\mathcal{E}_{x'}}$ also includes a series of depolarizing channel applied to the system qubits.

\section{Appendix: integrals and Euler constant}

The integral representation of the Euler constant is given by 
\begin{equation}
\gamma  =  - \int_0^\infty  {{e^{ - x}}\log x} dx = 0.57721 ...\ .
\end{equation}
A related integral is given by
\begin{equation}
\int_0^\infty  {{e^{ - x}}{{(\log x)}^2}} dx = {\gamma ^2} + \frac{{{\pi ^2}}}{6} \ .
\end{equation}
Recall that 
\begin{equation}
\sum\limits_{x,x'} {B{{\left( {x,x'} \right)}^2}}  = {\sum\limits_{x,x'} {\left| {\log {P_{{\text{qc}}}}\left( {x|{U_{x'}}} \right)} \right|} ^2} \ .
\end{equation}
To have an estimation of the value of left-hand side, we follow a similar procedure as in Eq.~(\ref{SP2IfP2oc}) and write,
\begin{equation}
{\sum\limits_x {\left| {\log {P_{{\text{qc}}}}\left( {x|{U_{x'}}} \right)} \right|} ^2} = \int_0^\infty  {f\left( P \right){{\left( {\log P} \right)}^2}dP} \ .
\end{equation}
We further assume that each of the quantum circuit $U_{x'}$ obeys Porter-Thomas distribution (see Eq.~(\ref{fPN2emNPPtd})), which means that the right-hand side becomes
\begin{equation}
{N^2}\int_0^\infty  {{e^{ - NP}}{{\left( {\log P} \right)}^2}dP} \ .
\end{equation}
Let us now consider the following integral, 
\begin{equation}
I_0 \equiv \int_0^\infty  {{e^{ - NP}}{{\left( {\log NP} \right)}^2}dNP} = {\gamma ^2} + \frac{{{\pi ^2}}}{6} \ ,
\end{equation}
which is equal to  
\begin{equation}
I_0 = \int_0^\infty  {{e^{ - NP}}{{\left( {\log N + \log P} \right)}^2}dNP} \ .
\end{equation}
This integral can be decomposed into three terms,
\begin{equation}\label{I0I1I2I3tic}
I_0 = I_1 + I_2 +I_3 \ ,
\end{equation}
where the first term on the right is given by,
\begin{equation}
{I_1} \equiv {(\log N)^2}\int_0^\infty  {{e^{ - NP}}dNP}  = {(\log N)^2} \ ,
\end{equation}
and the second term is given by 
\begin{equation}
{I_2} \equiv 2\log N\int_0^\infty  {{e^{ - NP}}\log P \, dNP} \ .
\end{equation}
To obtain $I_2$, let us now write the Euler constant as
\begin{equation}
 - \gamma  = \int_0^\infty  {{e^{ - NP}}(\log NP) dNP} \ ,
\end{equation}
where the right hand side can be written as 
\begin{equation}
\log N\int_0^\infty  {{e^{ - NP}}dNP}  + \int_0^\infty  {{e^{ - NP}}(\log P) \, dNP} \ .
\end{equation}
Consequently, we have 
\begin{equation}
 - \int_0^\infty  {{e^{ - NP}}(\log P) \, dNP}  = \log N + \gamma \ ,
\end{equation}
and hence 
\begin{equation}
{I_2} =  - 2\log N\left( {\log N + \gamma } \right) \ .
\end{equation}
Lastly, the third term on the right of Eq.~(\ref{I0I1I2I3tic}) is given by  
\begin{equation}
{I_3} \equiv \int_0^\infty  {{e^{ - NP}}{{(\log P)}^2}dNP} \ .
\end{equation}
Putting these together, ${I_3} = {I_0} - {I_1} - {I_2}$, and is given by
\begin{equation}
{I_3} = {\left( {\log N + \gamma } \right)^2} + \frac{{{\pi ^2}}}{6} \ .
\end{equation}

Note that we can also write
\begin{equation}
\sum\limits_x {{{\left| {\log {P_{{\text{qc}}}}\left( {x|{U_{x'}}} \right)} \right|}^2}}  = N{I_3} \ ,
\end{equation}
which gives the values for the bound for quantum circuits with Porter-Thomas distribution.



\end{document}